\documentclass[aps,prb,twocolumn,showpacs,amsmath]{revtex4}
\usepackage{latexsym}
\usepackage{amssymb}
\usepackage{graphicx}
\usepackage{bm}
\usepackage[english]{babel}
\usepackage[latin1]{inputenc}
\usepackage{amsmath}
\usepackage{amsfonts}

\newcommand{\pdag}{{\phantom{\dagger}}}
\newcommand{\bq}{\begin{equation}}
\newcommand{\eq}{\end{equation}}
\newcommand{\bn}{\begin{eqnarray}}
\newcommand{\en}{\end{eqnarray}}

\begin{document}

\title{Full counting statistics of a single-molecular quantum dot}

\author{Bing Dong}
\affiliation{Key Laboratory of Artificial Structures and Quantum Control (Ministry of Education), Department of Physics, Shanghai 
Jiaotong University, 800 Dongchuan Road, Shanghai 200240, China}

\author{G. H. Ding}
\affiliation{Key Laboratory of Artificial Structures and Quantum Control (Ministry of Education), Department of Physics, Shanghai 
Jiaotong University, 800 Dongchuan Road, Shanghai 200240, China}

\author{X. L. Lei}
\affiliation{Key Laboratory of Artificial Structures and Quantum Control (Ministry of Education), Department of Physics, Shanghai 
Jiaotong University, 800 Dongchuan Road, Shanghai 200240, China}

\begin{abstract}

We investigate the full counting statistics of a single quantum dot strongly coupled to a local phonon and weakly 
tunnel-connected to two metallic electrodes. By employing the generalized nonequilibrium Green function method and the 
Lang-Firsov transformation, we derive an explicit analytical formula for the cumulant generating function, which makes one to be 
able to identify distinctly the elastic and inelastic contributions to the current and zero-frequency shot noise. We find that at 
zero temperature, the inelastic effect causes upward steps in the current and downward jumps in the noise at the bias voltages 
corresponding to the opening of the inelastic channels, which are ascribed to the vibration-induced complex dependences of 
electronic self-energies on the energy and bias voltage. More interestingly, the Fano factor exhibits oscillatory behavior with 
increasing bias voltage and its minimum value is observed to be smaller than one half.    

\end{abstract}

\date{\today}


\maketitle

\section{Introduction}

Recent progress in nanotechnology has facilitated the fabrication of single-electron tunneling devices using organic molecules. A 
variety of intriguing effects has been observed in the transport properties of the single-molecule transistors (SMTs) due to the 
couplings between the intrinsic mechanical degree of freedom (phonon, vibron) in molecules and electrons during 
tunneling.\cite{hPark,jPark,Zhitenev,Weig,Yu,Pasupathy,LeRoy,Sapmaz} For instance, the obvious phonon-assisted current steps have 
been measured in the current-bias-voltage characteristic of a variety of individual molecules connected to metal 
electrodes.\cite{hPark,jPark,Zhitenev,Weig,Yu,Pasupathy} More interesting transport properties, e.g. the Franck-Condon blockade 
in the current steps and negative differential conductance due to nonequilibrated phonon excitation, have been experimentally 
demonstrated in the device of a suspended single-wall carbon nanotube.\cite{LeRoy,Sapmaz} 

These experimental observations have stimulated great interest in the theoretical investigations. In recent years, a variety of 
different approaches have been developed to study the transport properties and current fluctuation characteristics in the 
electron-phonon coupled systems, mainly containing the kinetic-equation approach (master 
equations),\cite{Bose,McCarthy,Mitra,Koch,Koch2,Zazunov,Shen,Dong2,Dong4,Avriller} the nonequilibrium Green function (NGF) 
techniques,\cite{Frederiksen,Galperin2,Viljas,Vega,Egger,Entin,Dash,Haule,Ness,Emberly,Dong1,Dong3,Lundin,Flensberg,Chen,Galperin
,Galperin3,Hartle,Zazunov2} and the diagrammatic Monte Carlo simulation.\cite{Muhlbacher} It is well-known that the NGF is a most 
powerful method to study nonequilibrium behavior of a many-body system. Within the NGF theoretical formulation, various 
self-consistent second-order perturbation calculations have been carried out, on the weak electron-phonon interaction (EPI) 
strength, to examine the inelastic correction to the nonlinear conductance of the 
SMTs.\cite{Frederiksen,Galperin2,Viljas,Vega,Egger,Entin,Dash} On the other hand, in order to study the strong EPI effects, two 
authors of this paper proposed a nonperturbative analysis of the inelastic effects on 
current\cite{Haule,Ness,Emberly,Dong1,Dong3} and its fluctuations\cite{Dong3} by mapping of the many-body EPI problem onto a 
multichannel single-electron scattering problem.\cite{Bonca} This mapping technique is only valid in the limit of weak electronic 
tunnel-couplings between the molecular quantum dot (QD) and electrodes since the so-called Fermi sea effect is neglected in the 
mapping procedure. To circumvent this drawback and to cover more wide ranges of system parameters, e.g. the EPI and 
tunnel-coupling strengths, another nonperturbative scheme has been recently developed based on the NGF in the polaron 
representation.\cite{Lundin,Flensberg,Chen,Galperin,Hartle,Zazunov2} In particular, Galperin {\it et.al.} formulated a fully 
self-consistent solution of both electronic and phononic GFs by employing equation of motion method to establish the Dyson-type 
coupled equations.\cite{Galperin} Later on, he further developed this approach to study the zero-frequency noise spectrum of 
SMTs.\cite{Galperin3} Because the Wick theorem can not be applied to calculate the current-current correlation function of the 
EPI system, he instead made use of the noise formula of the noninteracting system and simply replaced the electronic GFs in the 
noninteracting noise formula with the self-consistently calculated ones. Moreover, this approach was extended to consider the 
inelastic effect of multimode vibrational dynamics.\cite{Hartle}       

Nowadays, there is continually increasing interest in the full counting statistics (FCS) of charge transport in nanocale 
system.\cite{Nazarov,Blanter} This remarkable concept was first proposed by Levitov and Lesovik to describe the whole probability 
distribution of transmitted charge during a fixed time interval in a mesoscopic conductor.\cite{Levitov} It is therefore an 
intriguing task to examine the FCS of electronic tunneling in the presence of EPI. Employing the master equations, the inelastic 
effect on the FCS has been studied in the resonant tunneling regime.\cite{Koch2,Dong4,Avriller} For the phase-coherent transport 
through an interacting system, the NGF is required. 
Recently, Gogolin and Komnik have generalized the Meir-Wingreen NGF formulation for the quantum transport in mesoscopic system to 
the FCS issue, and derived a {\em generic expression} for the generating function of the cumulants expressed only in terms of the 
local Keldysh GFs of the central region, which is valid in any types of the central region, noninteracting or 
interacting.\cite{Gogolin} In this theory, it is no need to directly calculate the current-current correlation functions by 
employing the Feynman diagram technique. Instead a Schwinger external source, i.e. here a fictitious {\em measuring field} 
$\lambda$ in the tunneling Hamiltonian, is introduced to count the numbers of transmitted electrons and functional derivative is 
provoked at the end of calculation to generate the cumulants of charge current distribution.\cite{Levitov1} Another advantage of 
this Hamiltonian approach is that it can automatically contain the vertex corrections in the current-current correlation 
functions. With help of the generalized Schwinger-Keldysh GF technique, inelastic effects on the FCS in SMTs have been recently 
investigated, in which a compact analytic expression for the FCS was derived under the assumption that vibration mode is at 
equilibriated state.\cite{Schmidt} These authors focused their studies on the concrete behaviors of the current and shot noise 
jumps, upward or downward, due to phonon excitation when the first inelastic channel is opening. Remarkably, the negative 
contribution to noise due to vibration excitation has been experimentally observed by recent shot noise measurements on Au atomic 
contact,\cite{Kumar} and has been further confirmed by a subsequent calculation of the inelastic shot noise signals in Au and Pt 
atomic point contacts from first principles.\cite{Avriller2}  
Moreover, the effect of vibrational heating on FCS has been further considered,\cite{Urban,THPark} and analytical results on FCS 
accounting for nonequilibrium phonon distributions have been obtained.\cite{Utsumi} Nevertheless, all these studies employed the 
second-order perturbation expansion to evaluate the electronic and phononic self-energies, and consequently they are valid for 
the regime of weak EPI. The full knowledge of the inelastic effects on FCS in the regime of strong EPI is still less 
studied.\cite{Maier} This constitutes the purpose of the present paper.     

The rest of the paper is organized as follows. In Sec.~II, we introduce the model Hamiltonian of a molecular QD. In Sec.~III, we 
present the theoretical formulation for the FCS calculation in the presence of EPI. In particular, the explicit expressions of 
the FCS, current, and zero-frequency shot noise are derived. In Sec.~IV, we carry out numerical calculations of differential 
conductance, shot noise and Fano factor, and discuss these results. Finally, a brief summary is given in Sec.~V.

\section{Model}

In this paper, we consider a simple model for a molecular QD with one spinless level (electronic energy $\epsilon_d$) coupled to 
two electrodes left (L) and right (R) (each a freee electron reservoir at its own equilibrium), and also linearly coupled to a 
single vibrational mode (phonon) of the molecule having frequency $\omega_0$ with coupling strength $g_{ep}$. The model 
Hamiltonian is
\begin{subequations}
\bq
H = H_{leads}+H_{mol}+H_{T}, \label{ham}
\eq
with
\bn
H_{leads} &=& \sum_{\eta, {\bf k}} \varepsilon_{\eta {\bf k}} c_{\eta {\bf k}}^\dagger c_{\eta {\bf k}}^\pdag, \\
H_{mol} &=& \varepsilon_d d^{\dagger} d^\pdag + \omega_0 a^\dagger a + g_{ep} d^\dagger d^\pdag (a^\dagger + a), \label{Hmol} \\
H_T &=& \sum_{\eta,{\bf k}} (\gamma_{\eta} e^{-i\lambda_\eta(t)/2} c_{\eta {\bf k}}^\dagger d + {\rm H.c.}),
\en
\end{subequations}
where $c_{\eta{\bf k}}^\dagger$ ($c_{\eta{\bf k}}$) is the creation (annihilation) operator of an electron with momentum ${\bf 
k}$, and energy $\varepsilon_{\eta {\bf k}}$ in lead $\eta$ ($\eta=L,R$), and $d^\dagger$ ($d$) is the corresponding operator for 
a spinless electron in the QD. $a^\dagger$ ($a$) is phonon creation (annihilation) operators for the vibrational mode (energy 
quanta $\omega_0$). $\gamma_{\eta}$ describes the tunnel-coupling matrix element between the QD and lead $\eta$. The 
corresponding coupling strength is defined as $\Gamma_\eta = 2\pi \sum_{k} |\gamma_\eta|^2 \delta(\omega-\varepsilon_{\eta {\bf 
k}})$, which is assumed to be independent of energy in the wide band limit. In order to investigate the full counting statistics 
(FCS), an artificially measuring field $\lambda_\eta (t)$ is introduced with respect to the lead $\eta$ on the Keldysh contour: 
$\lambda_\eta(t)=\lambda_{\eta -} \theta(t) \theta({\cal T}-t)$ on the forward path and $\lambda_\eta(t)=\lambda_{\eta +} 
\theta(t) \theta({\cal T}-t)$ on the backward path (${\cal T}$ is the measuring time during which the counting fields are 
non-zero and the counting fields will be set to be opposite constants on the forward and backward Keldysh contour as 
$\lambda_{\eta -}=-\lambda_{\eta +}=\lambda_\eta$ in the final derivation).\cite{Levitov1,Gogolin} Throughout we will use natural 
units $e=\hbar=k_{\rm B}=1$.

For dealing with the problem involving strong electron-phonon interaction, it is very convenient to apply a standard Lang-Firsov 
canonical transformation, $S=g d^\dagger d (a^\dagger - a)$ ($g=g_{ep}/\omega_0$), to the Hamiltonian 
Eq.~(\ref{ham}),\cite{Mahan} leading to a transformed Hamiltonian
\begin{subequations}
\bn
\widetilde{H}&=&e^{S} H e^{-S} = H_{leads} + \widetilde{H}_{mol} +\widetilde{H}_{T}, \label{tranH}\\
\widetilde{H}_{mol} &=& \widetilde{\varepsilon}_d \widetilde{d}^\dagger \widetilde{d} + \omega_0 a^\dagger a = 
\widetilde{\varepsilon}_d d^\dagger d + \omega_0 a^\dagger a, \\
\widetilde{H}_{T} &=& \sum_{\eta,{\bf k}} (\gamma_{\eta} e^{-i\lambda_\eta(t)/2} c_{\eta {\bf k}}^\dagger \widetilde{d} + {\rm 
H.c.}) \cr
&=& \sum_{\eta,{\bf k}} (\gamma_{\eta} e^{-i\lambda_\eta(t)/2} c_{\eta {\bf k}}^\dagger d X + {\rm H.c.}). \label{tunneling}
\en
Here $\widetilde{\varepsilon}_d = \varepsilon_d - \frac{g_{ep}^2}{\omega_0}$ is the renormalized energy level of the QD and 
$\widetilde{d}=d X$ denotes the new Fermionic operator dressed by the phononic shift operator $X$, 
\bq
X = e^{g (a-a^\dagger)}.
\eq
\end{subequations}
Therefore, the transformed Hamiltonian is equivalent to a noninteracting resonant-level model with a vibration modified dot-lead 
tunneling described by the shift operator $X$ in Eq.~(\ref{tunneling}), which is responsible for the observation of the 
Franck-Condon steps in the current-voltage characteristics of the single molecular transistor. This noninteracting effective 
Hamiltonian $\widetilde{H}$ Eq.~(\ref{tranH}) is our starting point for the FCS investigation in the following section.

\section{Theoretical Methods}

\subsection{Adiabatic Potential for FCS}

To investigation the probability distribution $P_{q_L,q_R}$ of the charge $q_\eta$ to be transferred through the QD to lead 
$\eta$ during the measuring time, we should calculate the so-called cumulant generating function (CGF) 
$\chi(\lambda)\equiv\chi(\lambda_L,\lambda_R)=\sum_{q_L,q_R} P_{q_L,q_R} e^{i \sum_{\eta} q_\eta\lambda_\eta}$ for the 
two-terminal QD, which can be determined as a Keldysh partition function:\cite{Levitov1}
\bq
\chi(\lambda)= \left\langle T_{\cal C} e^{-i \int_{\cal C} \widetilde{H}_{T}(t) dt} \right\rangle_\lambda,
\eq
where $T_{\cal C}$ denotes time ordering along the Schwinger-Keldysh contour ${\cal C}$ and the expectation value is written in 
the interaction picture with respect to the effective Hamiltonian, $H_{leads}+\widetilde{H}_{mol}$. According to 
Ref.~\onlinecite{Gogolin}, to calculate the CGF $\chi(\lambda)$ it is technically more convenient employing the {\em adiabatic 
potential} method: $\ln \chi(\lambda)=-i{\cal T } {\cal U}(\lambda_-, \lambda_+)=-i{\cal T } {\cal U}(\lambda, -\lambda)$, where 
the adiabatic potential ${\cal U}(\lambda_-, \lambda_+)$ is defined due to the nonequilibrium Feynman-Hellmann theorem as
\begin{eqnarray}
{\partial {\cal U}(\lambda_-, \lambda_+)\over {\partial \lambda_{\eta -}}} &=& \left\langle {\partial 
\widetilde{H}_T(t)\over{\partial \lambda_{\eta -}}} \right\rangle_\lambda \nonumber\\
&=& -{i\over 2}\sum_{{\bf k}} \left\langle \gamma_\eta e^{-i\lambda_{\eta -}/2} c_{\eta {\bf k}}^\dagger \widetilde{d} - {\rm 
H.c.} \right\rangle_\lambda ,
\end{eqnarray}
with the notation
\bq
\langle \cdots \rangle_\lambda = \frac{1}{\chi(\lambda_-,\lambda_+)} \left\langle T_{\cal C} \cdots e^{-i \int_{\cal C} 
\widetilde{H}_{T}(t) dt} \right\rangle_0 .
\eq
The further evaluation of the adiabatic potential amounts to calculations of the mixed GFs, $G_{d\eta {\bf k}}(t,t')=-i\langle 
T_{\cal C} \widetilde{d}(t) c_{\eta {\bf k}}^\dagger(t') \rangle_\lambda$ and $G_{\eta {\bf k}d}(t,t')=-i\langle T_{\cal C}  
c_{\eta {\bf k}}(t) \widetilde{d}^\dagger (t')\rangle_\lambda$, as ($t^+=t+0^+$)
\begin{eqnarray}
{\partial {\cal U}(\lambda_-, \lambda_+)\over {\partial \lambda_{\eta -}}} &=& \frac{\gamma_\eta}{2} \sum_{{\bf k}} \left [ 
e^{-i\lambda_{\eta -}/2} G_{d \eta {\bf k}}^{--}(t,t^+) \right. \cr
&& \left. - e^{i\lambda_{\eta -}/2} G_{\eta {\bf k}d}^{--}(t,t^+) \right ]. \label{adp0}
\end{eqnarray}
Bearing in mind the facts that the transformed Hamiltonian is noninteracting one and the canonical transformation do not alter 
the canonical commutation relations between Fermionic operators, these mixed GFs can be cast into combinations of the 
contour-ordered GFs of the QD involving dressed electronic operators, $G_{d}(t,t')$, and bare lead GFs, $g_{\eta{\bf k}}(t,t')$,
\bn
G_{d\eta {\bf k}}(t,t') &=& \int_{\cal C} dt'' G_{d}(t,t'') \gamma_\eta e^{i\lambda_\eta(t'')/2} g_{\eta {\bf k}}(t'',t'),\cr
G_{\eta {\bf k}d}(t,t') &=& \int_{\cal C} dt'' g_{\eta {\bf k}}(t,t'') \gamma_\eta e^{-i\lambda_\eta(t'')/2} G_{d}(t'',t'), 
\nonumber
\en
with
\bn
G_{d}(t,t')&=& -i\left\langle T_{\cal C} \widetilde{d}(t) \widetilde{d}^\dagger (t') \right\rangle_\lambda \cr
&=& -i\left\langle T_{\cal C} d(t) X(t) X^\dagger(t') d^\dagger (t') \right\rangle_\lambda, \\
g_{\eta {\bf k}}(t,t') &=& -i\left\langle T_{\cal C} c_{\eta {\bf k}}(t) c_{\eta {\bf k}}^\dagger (t') \right\rangle_\lambda.
\en
Performing the Keldysh disentanglement and substituting the results back into Eq.~(\ref{adp0}) one obtains
\bn
{\partial {\cal U}(\lambda_-, \lambda_+)\over {\partial \lambda_{\eta -}}} &=& \sum_{\bf k}\frac{\gamma_\eta^2}{2} \int dt_1 
\left[ e^{-i \bar{\lambda}_\eta/2} G_{d}^{-+}(t,t_1) g_{\eta {\bf k}}^{+-}(t_1,t^+) \right. \cr
&& \left. -  e^{i\bar{\lambda}_\eta/2} g_{\eta {\bf k}}^{-+}(t,t_1) G_{d}^{+-}(t_1,t^+) \right ], \label{adp1}
\en
with $\bar{\lambda}_\eta= \lambda_{\eta-} - \lambda_{\eta +}$. 
It is noticed that the adiabatic potential Eq.~(\ref{adp1}) is exactly equivalent to that given by Maier in 
Ref.~\onlinecite{Maier}. Until now, all derivations are exact and what is done next is to calculate the dressed electronic GF 
$G_{d}^{\alpha\beta}(t,t')$ ($\alpha,\beta=+,-$). 

\subsection{Nonequilibrium GF approach for electron-phonon coupled system}
 
Following Galperin {\it et al.},\cite{Galperin} we can use the usual Born-Oppenheimer adiabatic approximation to decouple 
electron and phonon dynamics, which leads to a factorized form of the GF $G_d(t,t')$ as a product of a pure electronic part and a 
phononic part,\cite{Flensberg}
\bq
G_{d}^{\alpha\beta}(t,t') \approx G_{c}^{\alpha\beta}(t,t') K^{\alpha\beta}(t,t'), \label{deGF} 
\eq
where
\bn
G_{c}(t,t')&=& -i \left\langle T_{\cal C} d(t) d^\dagger (t') \right\rangle_\lambda, \\ 
K(t,t') &=& \left\langle T_{\cal C} X(t) X^\dagger(t') \right\rangle_\lambda.
\en
The corresponding Feynman diagram in perturbation theory is shown schematically in Fig.~1(a). This decoupling is valid in the 
limit of a weak molecule-lead tunnel-couplings implying a relatively long residence time of the electron on the molecule, i.e., 
$\Gamma_{\eta} \ll \omega_0$. 

Furthermore, we assume a extremely strong dissipation of the primary phonon mode to a thermal bath, e.g., to a substrate or a 
backgate. This means that the oscillator restores to its equilibrium state so quickly that it has no time to play a reaction to 
the electronic system when it is stimulated to an unequilibrated state by external-bias-voltage-driven tunneling electrons. In 
this situation, the oscillator can be described by an equilibrium Bose distribution $n_B=(e^{\omega_0/T}-1)^{-1}$ at the 
temperature $T$ and the phonon shift generator GF $K(t,t')$ can be replaced by its equilibrium correlation function,\cite{Mahan}
\bq
K(t,t')= \left (
\begin{array}{cc}
e^{-\phi(|\tau|)} &  e^{-\phi(\tau)} \\
e^{-\phi(-\tau)} &  e^{-\phi(-|\tau|)} \\
\end{array}
\right ),    
\eq
where $\phi(\tau)$ is defined as ($\tau=t-t'$) 
\bq
\phi(\tau) = g^2 \left [ n_B(1-e^{i\omega_0 \tau}) + (n_B+1) (1-e^{-i\omega_0 \tau}) \right ].
\eq
It is noted that in this approximation, the phononic GF $K(t,t')$ becomes irrespective of the counting field $\lambda$.

\begin{figure}[htb]
\includegraphics[height=8cm,width=8.5cm]{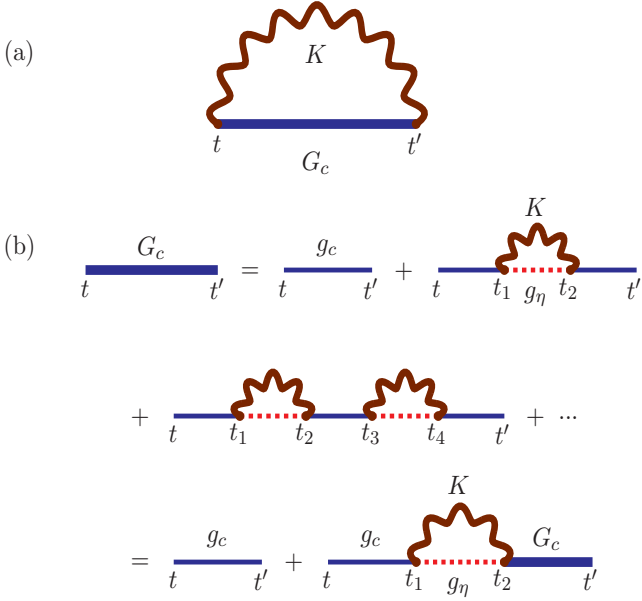}
\caption{(Colour online) The Feynman diagrams taken into account in our calculations for the EPI system. (a) The diagram for the 
factorized GF $G_d(t,t')$. The thick solid line denotes the pure electronic GF $G_c(t,t')$ and the wiggly line represents the 
phonon cloud propagator $K(t,t')$. (b) The set of Feynman diagrams and Dyson equation for the pure electronic GF $G_c(t,t')$. The 
thin solid line is the bare dot GF $g_c$, and the dashed-line denotes the GF $g_\eta$ ($\eta=L,R$) of the leads.}
\label{fig1}
\end{figure}

Therefore, the next step is to calculate the contour-ordered electronic GF of the QD, $G_{c}^{\alpha\beta}(t,t')$, based on the 
transformed Hamiltonian $\widetilde{H}$ Eq.~(\ref{tranH}). 
Nevertheless, since the transformed tunneling Hamiltonian Eq.~(\ref{tunneling}) involves the exponential operators $X$ and 
$X^\dagger$, one can not apply Wick's theorem to derive the Dyson-like equation of the pure electronic GF $G_c$. As an 
alternative method, an equation-of-motion (EOM) procedure has been usually used as approximations in 
literature.\cite{Flensberg,Galperin,Grover}
Here, using the transformed Hamiltonian $\widetilde{H}$, we derive the EOM for the contour-ordered electronic GF $G_{c}(t,t')$ as
\bn
&&\left ( i{\frac \partial {\partial t}}-\tilde\varepsilon_d \right )G_c(t,t')=\delta_{\cal C}(t-t') \nonumber\\
&&- \sum_{\eta {\bf k}} i\gamma_{\eta} e^{i\lambda_\eta(t)/2} \langle T_{\cal C}X^\dagger(t) c_{\eta {\bf k}}(t) d^\dagger(t') 
\rangle_{\lambda}. \label{gfct}
\en
Then we consider the EOM for the three-point GF $\langle  T_{\cal C}X^\dagger(t) c_{\eta {\bf k}}(t_1) d^\dagger(t') 
\rangle_{\lambda}$. It gives  
\bn
 &&\left ( i{\frac \partial {\partial t_1}}-\tilde\varepsilon_{\eta {\bf k}} \right )\langle  T_{\cal C}X^\dagger(t) c_{\eta {\bf 
k}}(t_1) d^\dagger(t')\rangle_{\lambda}  \nonumber\\
&&= \gamma_{\eta} e^{-i\lambda_\eta(t_1)/2} \langle T_{\cal C}X^\dagger(t) X(t_1)d(t_1)d^\dagger(t') \rangle_{\lambda}\;,
\en
or in the integration form as
\bn
&& \langle  T_{\cal C}X^\dagger(t) c_{\eta {\bf k}}(t_1) d^\dagger(t')\rangle_{\lambda} = \gamma_{\eta}\int_C dt_2 
e^{-i\lambda_\eta(t_2)/2} \cr
&& \times g_{\eta\bf k}(t_1,t_2)\langle T_{\cal C}X^\dagger(t) X(t_2)d(t_2)d^\dagger(t') \rangle_{\lambda}.
\en
By taking the time limit $t_1\rightarrow t$ in the above equation, and substitute it to Eq.~(\ref{gfct}), one can obtain the EOM 
for GF $G_c$ exactly as follows
\bn
&&\left ( i{\frac \partial {\partial t}}-\tilde\varepsilon_d \right )G_c(t,t')=\delta_{\cal C}(t-t')
- \sum_{\eta {\bf k}}\int_C dt_1 i\gamma_{\eta}^2 \nonumber\\
&& \times e^{i[\lambda_\eta(t)-\lambda_\eta(t_1)]/2}  g_{\eta\bf k}(t,t_1)\langle T_{\cal C}X^\dagger(t) 
X(t_1)d(t_1)d^\dagger(t')\rangle_{\lambda}.\nonumber\\
\en
Then we will make an approximation the same as in Eq.~(\ref{deGF}) to decompose the dressed propagator:
\bq
 \langle T_{\cal C} X^\dagger(t) X(t_1) d(t_1) d^\dagger(t')\rangle_{\lambda} \approx  i K(t_1,t) G_c(t_1,t'),
\eq
and consequently obtain the Dyson equation for $G_c$
\bq
\left ( i{\frac \partial {\partial t}}-\tilde\varepsilon_d \right )G_c(t,t')=\delta_{\cal C}(t-t')
+\int_C \Sigma_{c\lambda}(t,t_1)G_c(t_1,t')\;,
\eq
in which $\Sigma_{c\lambda}(t,t_1)$ is the contour-ordered electronic self-energies in the time domain, which includes all 
couplings of the electronic degrees of freedom on the QD with those in the electrodes and the vibrational mode, and the counting 
fields as well,
\bq
\Sigma_{c\lambda}^{\alpha\beta}(t,t_1)= \sum_{\eta {\bf k}} e^{i(\lambda_{\eta\alpha}- \lambda_{\eta\beta})/2} \gamma_\eta^2 
g_{\eta {\bf k}}^{\alpha\beta}(t,t_1) K^{\beta\alpha}(t_1,t).
\eq
The Dyson equation can also be written as an integration in terms of the pure electronic GF $G_{c}$,
\bq
G_{c}(t,t')=g_{c}(t,t')+ \int_{\cal C} dt_1dt_2 g_{c}(t,t_1) \Sigma_{c\lambda}(t_1,t_2) G_{c}(t_2,t'). \label{rgfc}
\eq
where $g_c(t,t')$ denotes the free electron GF for the dot without tunneling-coupling.
It is clear that the ensuring GF $G_c(t,t')$ corresponds to summing over all the diagrams as shown in Fig.~1(b). This means that 
the present method accounts the vibration-modified-effect on electronic tunneling processes by embedding the phononic propagator 
into the tunneling self-energies. While the polaron tunneling approximation (PTA) scheme developed in Ref.~\onlinecite{Maier} 
considers the vibrational effect only in the bare electronic GF, $g_c$, but remains the tunneling self-energies unmodified by 
phonon cloud (see the corresponding Feynman diagram, Fig.~3 in Ref.~\onlinecite{Maier}). On the other hand, our Dyson series for 
$G_c$ is also different from those of single particle approximation,\cite{Lundin,Flensberg} which performs the same factorization 
for the full GF $G_d$ as ours but take no account of the phonon cloud in the Dyson series for $G_c$. 

Now we accomplish our calculation for the pure electronic GF $G_c$. Projecting Eq.~(\ref{rgfc}) onto the real time axis according 
to Langreth analytical continuation rules, and then performing Fourier transformation of the resulting equations gives an 
explicit expression for the electronic GF $G_{c}(\omega)$ (Noting that the counting fields $\lambda_\eta(t)$ are taken to be 
opposite constants in time on the forward and backward Keldysh contour):
\begin{widetext}
\bq
G_{c}(\omega) = {1\over {\cal D}_{\lambda}(\omega)}  \left (
\begin{array}{cc}
\omega- \widetilde\epsilon_d + \Sigma_{c0}^{+-}(\omega)-\Sigma_{c}^r(\omega) & \Sigma_{c\lambda}^{-+}(\omega) \\
\Sigma_{c\lambda}^{+-}(\omega) & -[\omega-\widetilde\epsilon_d- \Sigma_{c0}^{-+}(\omega) - \Sigma_{c}^r(\omega)] \\
\end{array}
\right ), \label{gf}
\eq
with
\bn
{\cal D}_{\lambda}(\omega) &=& [\omega-\widetilde\epsilon_d - \Sigma_{c}^r(\omega)][\omega-\widetilde\epsilon_d - 
\Sigma_{c}^a(\omega)]+ \Gamma_L \Gamma_{R} \sum_{nm} w_n w_m \left \{ f_L(\omega+n\omega_0) [1-f_{R}(\omega-m\omega_0)] \left [ 
e^{i(\bar{\lambda}_L - \bar{\lambda}_{R})/2}-1 \right ] \right. \cr
&& \left. + f_R(\omega+n\omega_0) [1-f_{L}(\omega-m\omega_0)] \left [ e^{-i(\bar{\lambda}_L - \bar{\lambda}_{R})/2}-1 \right ] 
\right\},
\en
\end{widetext}
where the lesser and greater self-energies of the electron can be expressed in frequency domain as 
\bn
\Sigma_{c\lambda}^{-+}(\omega) &=& \sum_{n=-\infty}^{\infty} w_n \Sigma_{c\lambda}^{(0),-+}(\omega+n\omega_0),\\
\Sigma_{c\lambda}^{+-}(\omega) &=& \sum_{n=-\infty}^{\infty} w_n \Sigma_{c\lambda}^{(0),+-}(\omega-n\omega_0), \\
\Sigma_{c0}^{\pm\mp}(\omega) &=& \Sigma_{c\lambda}^{\pm\mp}(\omega) \mid_{\lambda=0},
\en
and
\bn
\Sigma_{c\lambda}^{(0),-+}(\omega) &=& i\sum_{\eta} e^{i\bar{\lambda}_\eta/2} \Gamma_{\eta} f_{\eta}(\omega) ,\\
\Sigma_{c\lambda}^{(0),+-}(\omega) &=& -i\sum_{\eta} e^{-i\bar{\lambda}_\eta/2} \Gamma_{\eta} [1-f_{\eta}(\omega)].
\en
Here $f_\eta =[1+\exp {(\omega-\mu_\eta)/T}]^{-1}$ is the Fermi distribution function at temperature $T$ and chemical potential 
$\mu_{\eta}=E_F+V_{\eta}$ of lead $\eta$ ($E_F$ is the Fermi energy and $V_{\eta}$ is the bias-voltage applied to lead $\eta$). 
The factor $w_n$ is the weighting factor describing the electronic tunneling involving absorption or emission of $n$ phonons. At 
a finite temperature, 
\bq
w_n=e^{-g^2(2N_B+1)} e^{n\omega_0/2T} I_n(2g^2\sqrt{n_B(n_B+1)}),
\eq
where $I_n(x)$ is the $n$th Bessel function of complex argument. Moreover, the retarded self-energy in time domain can be defined 
in the usual way from the lesser and greater counterparts, $\Sigma_{c}^r(\tau)=\theta(\tau) [ \Sigma_{c0}^{+-}(\tau) - 
\Sigma_{c0}^{-+}(\tau)]$, and thus its expression in frequency domain is
\bn
\Sigma_{c}^r(\omega) &=& \sum_{\eta n} w_n \int \frac{d\omega'}{2\pi} \left \{ \frac{\Gamma_\eta f_\eta(\omega')}{\omega + 
n\omega_0 - \omega' +i0^+} \right. \cr
&& \left. + \frac{\Gamma_\eta [1-f_\eta(\omega')]}{\omega - n\omega_0 -\omega' +i0^+} \right \}. \label{selfenergy}
\en
It is observed that the vibration-modified electronic self-energy due to tunneling is highly dependent on the applied bias 
voltage as shown in Fig.~\ref{fig2} in the following section, in contrast to the noninteracting QD-lead system where the 
tunneling induced self-energy is assumed to be a constant, $\Sigma^{r}(\omega)=-i(\Gamma_L+\Gamma_R)/2$, in the wide band limit. 
Finally, for the purpose of analyzing the nonlinear transport properties, one needs calculate the local spectral function of the 
central region, which can be defined as
\bn
A(\omega) &=& -i[ G_d^{+-}(\omega) - G_d^{-+}(\omega)] {\big |}_{\lambda=0}\cr
&=& -i\sum_{n} w_n [ G_{c}^{+-}(\omega-n\omega_0) - G_{c}^{-+}(\omega+n\omega_0)] {\big |}_{\lambda=0}. \cr
&& \label{dos}
\en
 
\subsection{Expressions for FCS, Current, and Shot Noise}

Inserting all these results derived in above subsection into Eq.~(\ref{adp1}) and integrating over $\lambda_{\eta-}$ and setting 
$\lambda_{\eta-}=-\lambda_{\eta+}=\lambda_{\eta}$, we can yield an explicit analytical formula for the CGF of the electronic 
transport through a single molecular QD in presence of strong electron-phonon interaction   
\bn
\ln \chi(\lambda) &=& {\cal T} \int {d\omega\over{2\pi}} \ln \left \{ 1 + \sum_{nm} T_{nm}(\omega) \right. \cr
&& \times \left [ f_{L+n} (1-f_{R-m}) \left ( e^{i \lambda} -1 \right ) \right. \cr
&& \left.\left. + f_{R+m} (1-f_{L-n}) \left ( e^{-i \lambda} -1 \right ) \right ] \right \}, \label{cgf} 
\en
where $T_{nm}(\omega)$ is the transmission coefficient of electron between the left and right electrodes involving vibrational 
quanta $n$ and $m$:
\begin{equation}
T_{nm}(\omega) = {\Gamma_L\Gamma_{R}w_n w_m \over {\cal D}_{0}(\omega)},
\end{equation}
with $\lambda\equiv \lambda_L - \lambda_R$, ${\cal D}_0(\omega)={\cal D}_{\lambda}(\omega)\mid_{\lambda=0}$ and $f_{\eta\pm n}$ 
is a shorthand for $f_{\eta}(\omega\pm n\omega_0)$.

It is known that one of the advantages of the FCS conception in quantum transport is that the FCS expression can be used to 
distinguish the elementary events of electronic tunneling, thus provide some insight into the relevant transport 
properties.\cite{Tobiska} Therefore, we can conclude from Eq.~(\ref{cgf}) that under the condition of weak tunneling and strong 
EPI, electronic transport through a molecular QD can still be regarded as three distinct independent processes: (i) electrons 
transmitted from the left electrode to the right with probability $P_+= \sum_{nm} T_{nm} f_{L+n}(1-f_{R-m})$; (ii) transmission 
from right to left with $P_-= \sum_{nm} T_{mn} f_{R+n}(1-f_{L-m})$; (iii) no transmission with $P_0=1-P_+-P_-$. Accordingly, the 
generating function for each process is $\chi=\sum_{\xi=+,-,0} P_{\xi} X_{\xi}$ with $X_\xi=e^{i\xi\lambda}$.
It is worth to notice that these transmission processes involve all possible phonon-assisted events. For example, the independent 
process (i) describes the specific electronic tunneling that an electron with energy $\omega$ in the left lead absorbs $n$ (if 
$n\geq 0$) or emits $n$ (if $n<0$) phonon in the left bridge, and tunnels through the central region, and eventually enters into 
the right lead with emitting $m$ (if $m\geq 0$) or absorbing $m$ (if $m<0$) phonon in the right bridge.  
Bearing in mind of these considerations, it can be addressed that the present FCS formula Eq.~(\ref{cgf}) is a direct extension 
of the original Levitov-Lesovik formula,\cite{Levitov} 
\bn
\ln \chi(\lambda) &=& {\cal T} \int {d\omega\over{2\pi}} \ln \left \{ 1 + T(\omega) \left [ f_{L} (1-f_{R}) \left ( e^{i \lambda} 
-1 \right ) \right. \right. \cr
&& \left.\left. + f_{R} (1-f_{L}) \left ( e^{-i \lambda} -1 \right ) \right ] \right \}, \label{LL} 
\en
to the inelastic electron transfer processes with either absorption or emission of phonon.    

Noticing the relation $w_{-n}=e^{-n \omega_0/T} w_n$, we can further deduce from Eq.~(\ref{cgf}) that in the present 
approximation, the FCS cumulants obey a universal relation 
\bq
\chi(V, \lambda)=\chi(V, -\lambda+i V/T),
\eq
which means that the detailed balance condition between the probabilities of opposite number of particles transferred through the 
QD remain valid even in the presence of electron-vibration interaction.\cite{Tobiska,Forster} The out-of-equilibrium fluctuation 
relations relate current correlation functions at any order at equilibrium to response coefficients of current cumulants of lower 
order.\cite{Tobiska,Forster}   

Based on the explicit analytical expression Eq.~(\ref{cgf}) of CGF, one can obtain all cumulants of charge transfer distribution 
through the molecular QD. We will however focus on the investigation of the first two cumulants, i.e., the average current 
through the system and the zero-frequency shot noise, in this paper, because they are the most easily accessible quantities in 
the experimental measurements. In specific, the average current $I$ from the left lead to the QD is evaluated as follows:
\bn
I &=& {2e\over \hbar} {1 \over {\cal T}} {\partial \ln \chi(\lambda)\over {\partial (i\lambda_L)}} {\bigg |}_{\lambda=0} 
= {2e\over h} \int d\omega  \sum_{nm} T_{nm}(\omega)  \cr
&& \times \left [ f_{L+n} (1-f_{R-m}) - f_{R+m} (1-f_{L-n}) \right ]. \label{current}
\en
From Eq.~(\ref{current}) the current can be separated as two contributions of elastic and inelastic parts, $I=I_{el}+I_{in}$, 
where the elastic current is 
\bq
I_{el} = {2e\Gamma_L \Gamma_R\over h} \int d\omega  \frac{w_0^2}{{\cal D}_0(\omega)} \left [ f_{L}(\omega)  - f_{R}(\omega) 
\right ]. \label{currentel}
\eq
While the zero-frequency current fluctuation $S$ is give by:
\begin{widetext}
\begin{eqnarray}
S = {4e^2\over \hbar}{1 \over {\cal T}}{\partial^2 \ln \chi(\lambda) \over {\partial (i\lambda_L)^2}} {\bigg |}_{\lambda=0} &=& 
{4e^2\over h} \int d\omega \left ( \sum_{nm} T_{nm}(\omega) [ f_{L+n} (1-f_{R-m}) + f_{R+m} (1-f_{L-n}) ] \right. \cr
&& \left. - \left \{ \sum_{nm} T_{nm}(\omega) [ f_{L+n} (1-f_{R-m}) - f_{R+m} (1-f_{L-n}) ] \right \}^2 \right ). \label{automf}
\en
\end{widetext}

Before ending this section, we compare our CGF formula Eq.~(\ref{cgf}) with the previous PTA result, Eq.~(2) in 
Ref.~\onlinecite{Maier}, which is obtained under the same limitation conditions, strong EPI and weak tunnel-coupling. As 
mentioned above, the PTA scheme takes no account of vibrational effect in the tunneling self-energy in its Dyson equation for 
calculating the full electronic GF $G_d$. One can argue that the PTA only considers virtual excitation of phonon in each 
electronic tunneling process, i.e., when an electron tunnels onto the molecule it excites the local phonon and fully de-excites 
the phonon upon leaving the dot. Therefore the electron after tunneling has the same energy as that before tunneling. This is why 
the PTA CGF [Eq.~(22) in Ref.~\onlinecite{Maier}] has the similar form with the original Levitov-Lesovik formula Eq.~(\ref{LL}).
While in the present approximation, after an electron tunnels into the molecular QD and excites the phonon, a virtual tunneling 
of electron into the leads is considered leading to excitations of particle-hole pairs in the leads. Then the electron tunnels 
out of the molecular QD and de-excites the phonon, but some particle-hole pairs remain in the leads, therefore energies of the 
electron before and after tunneling can be different. Physically, our results seem more reasonable because elastic and inelastic 
tunneling processes are both considered while only elastic tunneling processes are considered in the PTA scheme.

\section{Results and Discussions}

Here we carry out the numerical calculation of the current and zero-frequency shot noise through a single-molecular QD using 
Eqs.~(\ref{current}) and (\ref{automf}). For simplicity, we consider the system with symmetric tunnel-couplings to the leads, 
$\Gamma_L=\Gamma_R=0.1\omega_0$, $\Gamma=\Gamma_L+ \Gamma_R$, and assume the bias voltage is applied symmetrically, i.e., 
$\mu_{L/R}=\mu \pm V/2$. Therefore we can only consider positive bias voltage $V\geq 0$ in the following calculations. We also 
set the phonon energy $\omega_{0}=1$ as the unit of energy throughout the rest of the paper and choose the Fermi levels of the 
two leads as the reference of energy $\mu_{L}=\mu_{R}=\mu=0$ at equilibrium. The normalized EPI constant is set to be $g=1$ to 
ensure the validate of the approximation scheme involved in the present paper.

Below we mainly consider zero temperature, at which the weighting factor becomes
\bq
w_n= \left \{
\begin{array}{ll}
e^{-g^2}g^{2n}/n!, & n \geq 0, \\
0, & n<0,
\end{array}
\right.
\eq
meaning that only phonon emission processes are allowed. 

\subsection{Self energy and spectral function}

\begin{figure}[htb]
\includegraphics[height=7cm,width=9cm]{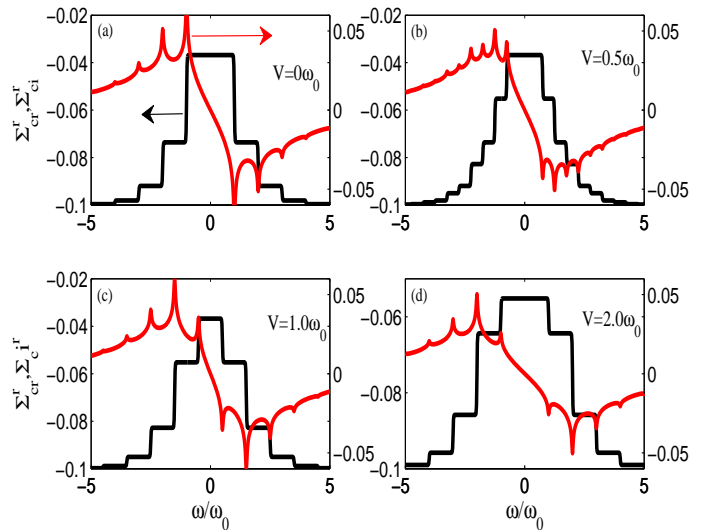}
\caption{(Colour online) The real part (red line) and imaginary part (black line) of the vibration-modified retarded 
self-energies are plotted for different bias-voltages, $V=0$ (a), $0.5\omega_0$ (b), $1.0\omega_0$ (c), and $2.0\omega_0$ (d), 
respectively, at zero temperature. The parameters used for calculation are taken as: $\Gamma_L=\Gamma_R=0.1\omega_0$, $g=1.0$.}
\label{fig2}
\end{figure}

We first examine the dependence of the tunneling-induced electronic self-energy, Eq.~(\ref{selfenergy}), on the bias voltage in 
Fig.~\ref{fig2} at zero temperature. We find that its imaginary part has explicit stepwise structures in frequency domain related 
to the opening of the inelastic channels, and the widths and heights of these steps are controlled by external applied bias 
voltage. Correspondingly, the real part of the self-energy shows multi-peaks with logarithmic singularities due to the 
Kramers-Kronig relations, which can be traced back to a previous work on EPI system by Engelsberg and Schrieffer for bulk 
Einstein phonons in 1963.\cite{Engelsberg} It is observed that the real parts of the self-energies, i.e., the values of the 
energy shift, are relatively small in the case of weak tunnel-coupling. 

\begin{figure}[htb]
\includegraphics[height=6cm,width=8cm]{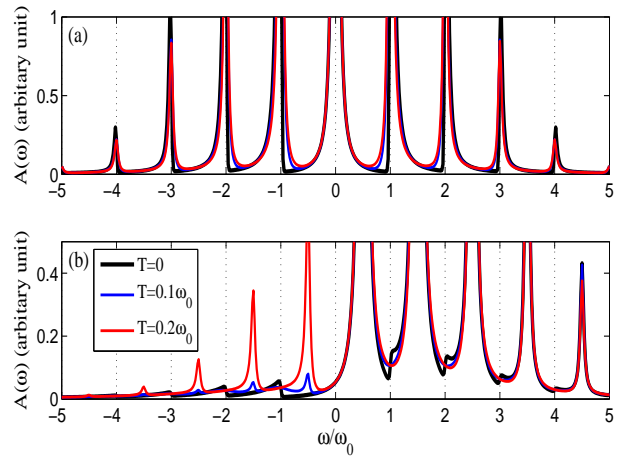}
\caption{(Colour online) The equilibrium calculated spectral function of the QD as a function of the energy $\omega$ for (a) 
$\widetilde{\varepsilon}_d=0$ and (b) $\widetilde{\varepsilon}_d=0.5\omega_0$ at different temperatures $T=0$, $0.1\omega_0$, and 
$0.2\omega_0$, respectively. The remaining parameters are the same as those in Fig.~\ref{fig2}.}
\label{fig3}
\end{figure}

We then calculate the equilibrium spectral function Eq.~(\ref{dos}) for the systems with $\widetilde{\varepsilon}_d=0$ and 
$0.5\omega_0$. As shown in Fig.~\ref{fig3}, one can find that the main effects of the electron-phonon coupling is the appearance 
of the phonon-assisted side peaks in the spectral function. At the zero temperature case and the renormalized level 
$\widetilde{\varepsilon}_d=0$, the main resonant peak at $\omega=0$ is Lorentzian in shape, while the phonon side peaks exhibit 
non-Lorentzian form due to stepwise jumps in the imaginary part of the self-energy as depicted in Fig.~\ref{fig2}(a). Peculiarly, 
these phonon side peaks symmetrically distribute in both sides of the energy axes at $\omega=\pm |n|\omega_0$ with gradually 
reduced heights. This behavior can be understood from the local spectral function Eq.~(\ref{dos}).\cite{Chen} At zero 
temperature, the local spectral function has two contributions, the lesser GF $G_c^{-+}(\omega+n\omega_0)$ and the greater GF 
$G_c^{+-}(\omega-n\omega_0)$ at $n\geq 0$. These two GFs correspond to the local electron and hole propagators, respectively, and 
thus are proportional to the occupation number $n_d$ for the QD electron or $1-n_d$ for the hole. For the system with 
$\widetilde{\varepsilon}_d=0$ and symmetrical tunnel-couplings to electrodes $\Gamma_L=\Gamma_R$, the QD is partially occupied by 
electrons, $n_d=1/2$. One can therefore interpret that the phonon side peaks at negative $\omega$ region result from the phonon 
emission by local electrons while the phonon side peaks at positive $\omega$ region originate from the phonon emission by local 
holes.   

When $\widetilde{\varepsilon}_d$ is far away from the the chemical potentials $\mu_L=\mu_R=\mu=0$, the side peaks become 
asymmetry on the two sides of the main peak located at $\omega=\widetilde{\varepsilon}_d$. For example, the spectral function of 
the system with $\widetilde{\varepsilon}_d=0.5\omega_0$ exhibits Lorentzian-type phonon side peaks only at positive $\omega$ 
region, $\omega=\widetilde{\varepsilon}_d+|n|\omega_0$, but no phonon side peak at negative $\omega$ region, because no electron 
occupies the QD, $n_d\simeq 0$. More interestingly, a small abrupt jump in the spectral function survives at 
$\omega=\pm|n|\omega_0$ as depicted in Fig.~\ref{fig3}(b), which is also stemming from the stepwise jumps in the imaginary part 
of the self-energy occurring at these frequencies corresponding to the opening of inelastic scattering processes. It is not 
surprise that with raising temperature $T$, all these novel features in the spectral function are gradually smoothed away. 
Besides, several phonon side peaks reemerge in negative energy regions due to the opening of phonon absorption channels at higher 
temperature [Fig.~\ref{fig3}(b)]. Furthermore, application of external bias voltage will change the occupation number of 
electrons at the QD, and will inevitably change the spectral function.    
It will be shown below that it is the complex dependences of the self-energy on the bias voltage in conjunction with the tiny  
features in the spectral function $A(\omega)$ that determines exotic properties of the nonlinear conductance and shot noise. 

\subsection{Current and differential conductance}

Before investigating nonlinear transport, we consider the zero-temperature linear conductance at first. It is easy from 
Eq.~(\ref{current}) to yield
\bq
G=\frac{dI}{dV}{\bigg |}_{V=0}=\frac{\Gamma_L \Gamma_R w_0^2}{[\widetilde{\varepsilon}_d+\Sigma_{cr}^r(0)]^2+ 
|\Sigma_{ci}^r(0)|^2},
\eq
with $\Sigma_{cr}^r(0)=0$ and $\Sigma_{ci}^r(0)=-i(\Gamma_L+\Gamma_R)w_0/2$. Therefore, in the linear transport regime, the 
effect of the strong electron-phonon interaction is just to narrow the resonance peak of the conductance due to the Franck-Condon 
blockade. Besides, the linear conductance exhibits no phonon sidebands as a function of the gate voltage. These two aspects of 
the linear conductance are in good agreement with the previous results for weak electron-phonon coupling systems based on the 
perturbative calculation up to the second-order of the electron-phonon coupling constant, $g^2$.\cite{Entin}

\begin{figure}[htb]
\includegraphics[height=5cm,width=8.5cm]{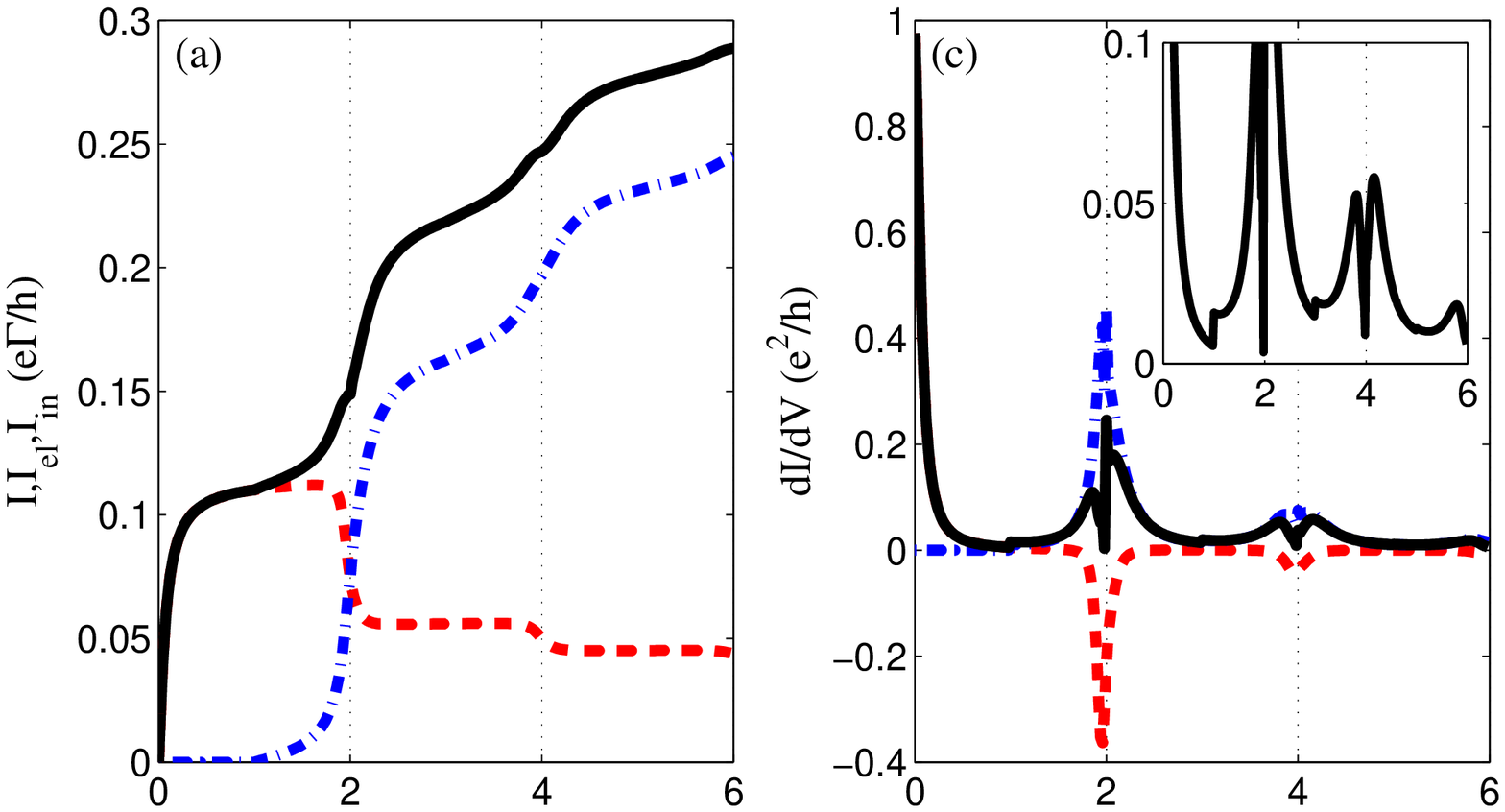}
\vspace{3mm}

\includegraphics[height=5.5cm,width=8.5cm]{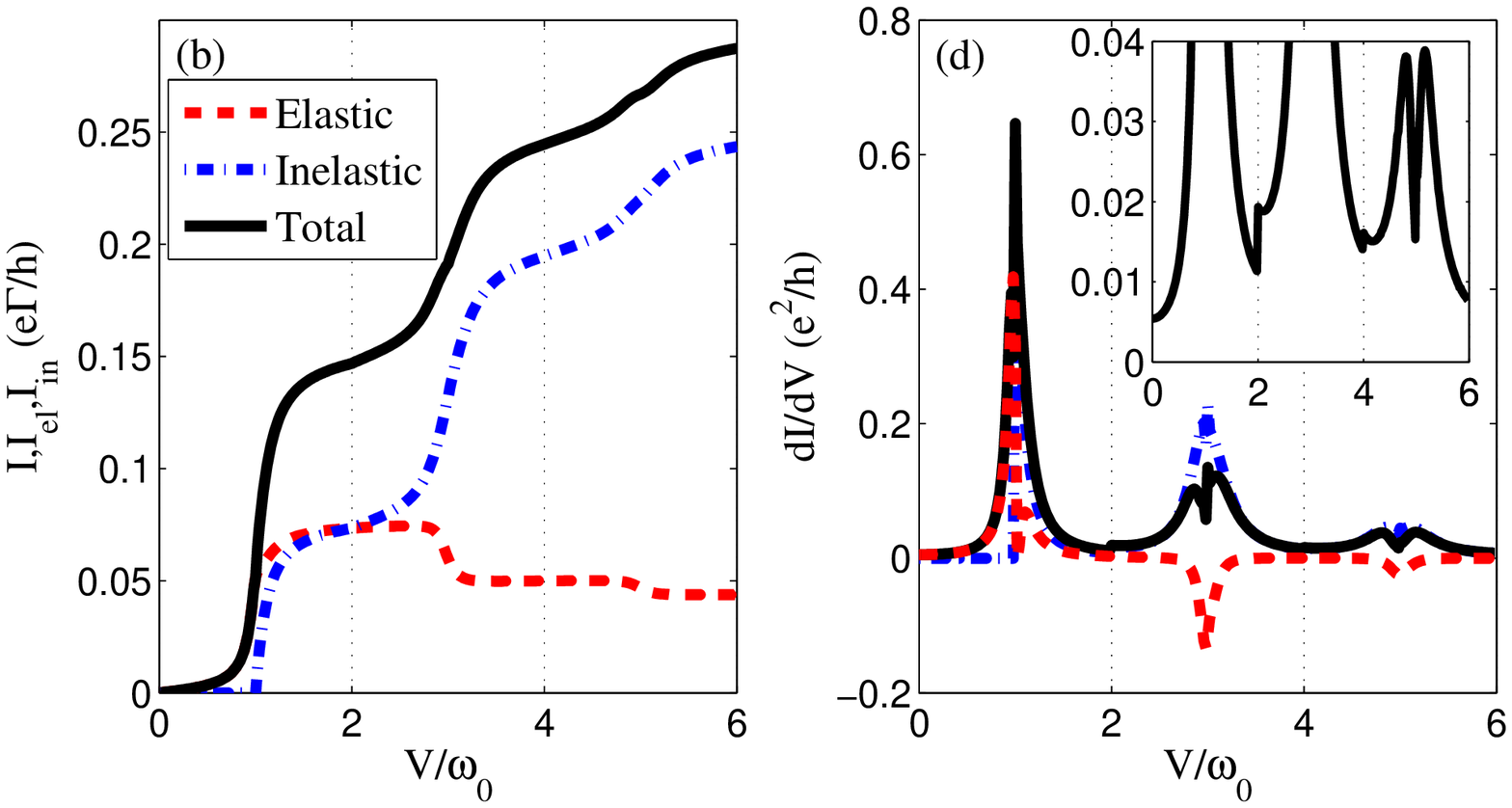}
\caption{(Colour online) (a, b) The calculated total current (solid line), elastic current (dashed line), inelastic current 
(dotted-dashed line); and (c, d) the corresponding differential conductances as functions of bias voltage for a single-molecular 
QD with $\widetilde{\varepsilon}_d=0$ (a, c) and $0.5\omega_0$ (b, d), respectively, at zero temperature. The remaining 
parameters are the same as those in Fig.~\ref{fig2}. The inset shows the enlarged tiny features of the respective differential 
conductances. (See text for details).}
\label{fig4}
\end{figure}

The situation is very different for the nonlinear transport as shown in Fig.~\ref{fig4}, in which we plot the currents $I$ and 
corresponding differential conductances $dI/dV$ as functions of bias voltage $V>0$ for the systems with 
$\widetilde{\varepsilon}_d=0$ and $0.5\omega_0$ at zero temperature. For the purpose of analysis, we also plot their 
corresponding elastic and inelastic parts. 
It is easy to obtain from Eq.~(\ref{current}) that only when the bias voltage is larger than the phonon energy, $V\geq\omega_0$, 
the inelastic current channels are opening, which leads to abrupt upward jumps of the differential conductance at $V=n \omega_0$ 
($n> 0$). Nevertheless, these upward jumps can be divided into two sorts, big jumps and tiny jumps. To obtain clear 
interpretation of these jumps, we give an explicit expression of the main contributive terms of the inelastic current at zero 
temperature as:   
\bq
I_{in}\simeq \frac{2e}{h}\Gamma_L \Gamma_R w_0 w_1 \int_{\omega_0- V/2}^{V/2} d\omega \left [\frac{1}{{\cal D}_0(\omega)} + 
\frac{1}{{\cal D}_0(-\omega)} \right ]. \label{currentin}
\eq
For the partially filled QD ($\widetilde{\varepsilon}_d=0$), the external bias voltage $V=\omega_0$ causes only a tiny jump due 
to the nonzero value of ${\cal D}_0(\pm V/2)\simeq (\omega_0/2)^2+|\Sigma_{ci}^r(\pm \omega_0/2)|^2$, but the bias voltage 
$V=2\omega_0$ results in a big jump owing to the minimum value in ${\cal D}_0(\pm \omega_0 \mp V/2)\simeq |\Sigma_{ci}^r(0)|^2$;  
while for the empty QD ($\widetilde{\varepsilon}_d=0.5\omega_0$), big jumps will occur at $V=\omega_0$ and $3\omega_0$ because of 
${\cal D}_0(\omega)\simeq (\omega-\omega_0/2)^2+|\Sigma_{ci}^r(\omega)|^2$. The tiny jumps at $V=2\omega_0$ and $4\omega_0$ are 
the remaining effect of the small abrupt jump in the spectral function as shown in Fig.~\ref{fig3}(b). 

Now we turn to discuss the elastic part of the tunneling current. The elastic current formula Eq.~(\ref{currentel}) can be 
simplified at zero temperature as 
\bq
I_{el}=\frac{2e}{h}\Gamma_L \Gamma_R w_0^2 \int_{- V/2}^{V/2} d\omega \frac{1}{{\cal D}_0(\omega)}. \label{currentel2}
\eq
As usually, the elastic current rises monotonously as the bias voltage of the left lead is increasing up to the energy level of 
the QD, $V=2\widetilde{\varepsilon}_d=0$ or $1.0\omega_0$, i.e., the resonant tunneling condition is reached. It is quite 
surprise, however, that the elastic current exhibits decrease steps with increasing further the bias voltages. To give an 
underlying interpretation of this decrease, we examine the derivative of the elastic current with respect to the bias voltage. 
Differentiating Eq.~(\ref{currentel2}) with respect to $V$, the nonlinear conductance can be written as 
$g^{el}=dI_{el}/dV=g_{1}^{el}+g_{2}^{el}$, with
\bq
g_{1}^{el}= \frac{e^2}{h}\Gamma_L \Gamma_R w_0^2 \left [ \frac{1}{{\cal D}_0(-V/2)} + \frac{1}{{\cal D}_0(V/2)} \right ],
\eq
and
\bn
g_{2}^{el}&\simeq & -\frac{2e^2}{h}\Gamma_L \Gamma_R w_0^2 \sum_{n=1}^{\infty} w_n \int_{-V/2}^{V/2} d\omega \frac{|\Sigma_{ci}^r 
(\omega)|}{{\cal D}_0^2 (\omega)} \cr
&& \times \left [ \Gamma_L \delta(\omega - n\omega_0 +V/2) + \Gamma_R \delta(\omega +n\omega_0 -V/2) \right ]. \cr
&&
\en
The first term, $g_{1}^{el}$, is proportional to the transmission probability $T_{00}(V/2)$ and results in the first resonant 
peak at $V=2\widetilde{\varepsilon}_d$; while the second term, $g_{2}^{el}$, is stemming from the bias-voltage-dependent 
self-energy and it always makes negative contribution and becomes predominant over the first term at $V=2n\omega_0+ 
2\widetilde{\varepsilon}_d$ ($n>0$) and at $V=2n\omega_0- 2\widetilde{\varepsilon}_d$ ($n>1$), which is responsible for decrease 
steps in the elastic current and the double-peak structure in the total differential conductance at $V=2\omega_0$, $4\omega_0$ 
for the QD with $\widetilde{\varepsilon}_d=0$ or at $V=3\omega_0$, $5\omega_0$ for $\widetilde{\varepsilon}_d=0.5\omega_0$. 

It should be noted that the inelastic scattering induced discontinuities, i.e., downward or upward steps, in the differential 
conductance have been previously reported based on the self-consistent Born approximation and the second-order perturbation 
calculations in the case of weak EPI.\cite{Galperin2,Egger,Entin} Our present nonperturbative calculations show more complex 
behavior for the systems with strong EPI and hard phonon $\omega_0\gg \Gamma$: tiny upward steps and double-peak profiles.      
 
All these tiny features in the differential conductance will be inevitably smeared away with increasing temperature, but those 
big jumps will survive (not shown here). Therefore, the differential conductance will still reflect the main profile of the 
spectral function of the molecular QD as shown in Fig.~\ref{fig3} at relatively high temperature. Besides, it is observed that 
the magnitudes of these big jumps gradually decrease with increasing bias voltage due to Franck-Condon blockade. 

\begin{figure}[htb]
\includegraphics[height=6cm,width=8.5cm]{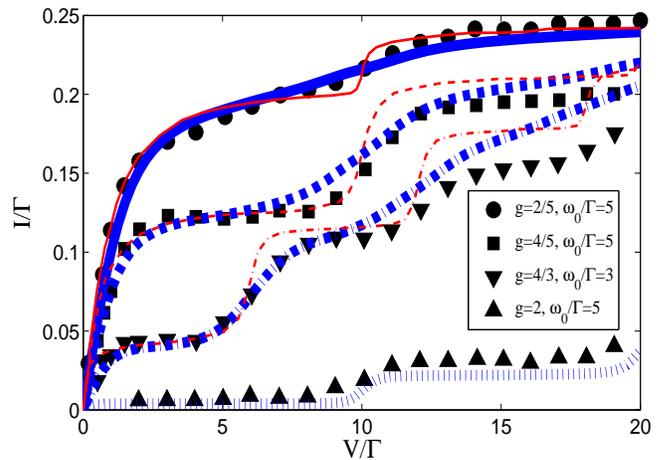}
\caption{(Colour online) The calculated current as functions of bias voltage for $\widetilde{\varepsilon}_d=0$ at the temperature 
$T=0.2\Gamma$. The thick-blue lines denote the present calculations, the thin-red lines are those of the PTA, and the discrete 
symbols represent the diagrammatic Monte Carlo data. Circles stand for the QD with $g=2/5$, $\omega_0/\Gamma=5$; squares for 
$g=4/5$, $\omega_0/\Gamma=5$; downward triangles for $g=4/3$, $\omega_0/\Gamma=3$; and upward triangles for $g=2$, 
$\omega_0/\Gamma=5$.}
\label{fig5}
\end{figure}

Before turning to discuss the shot noise, in order to estimate the quality of the present approximation, we compare our results 
with those of accurate diagrammatic Monte Carlo simulation,\cite{Muhlbacher} by plotting the calculated $I$-$V$ characteristics 
for several different molecular QD systems at a finite temperature $T=0.2\Gamma$, as shown in Fig.~\ref{fig5}. For comparison, we 
also plot the results calculated using PTA. It is clear that in the regime of moderate to large bias voltage $V$, our method 
exhibits better consistency with the Monte Carlo simulation than the PTA.

\subsection{Zero-frequency shot noise}

In what follows, we analyze the zero-frequency shot noise at zero temperature, which can be calculated using a simplified 
expression according to the Eq.~(\ref{automf})  
\bn
S &=& 2eI- \frac{4e^2}{h}(\Gamma_L \Gamma_R)^2 \sum_{nmn'm'} w_n w_m w_{n'} w_{m'} \cr
&& \times \int_{\omega_1}^{\omega_2} d\omega \frac{1}{{\cal D}_0^2(\omega)}, \label{s0}
\en 
with $\omega_1=\max(n\omega_0-V/2,n'\omega_0-V/2)$ and $\omega_2=\min(V/2-m\omega_0,V/2-m'\omega_0)$. We can also separate the 
shot noise as two contributions of elastic and inelastic parts, $S=S_{el}+S_{in}$, with the elastic part being
\bq
S_{el}=2eI_{el}- \frac{4e^2}{h}(\Gamma_L \Gamma_R)^2 w_0^4 \int_{-V/2}^{V/2} d\omega \frac{1}{{\cal D}_0^2(\omega)}.
\eq

\begin{figure}[htb]
\includegraphics[height=5cm,width=8.5cm]{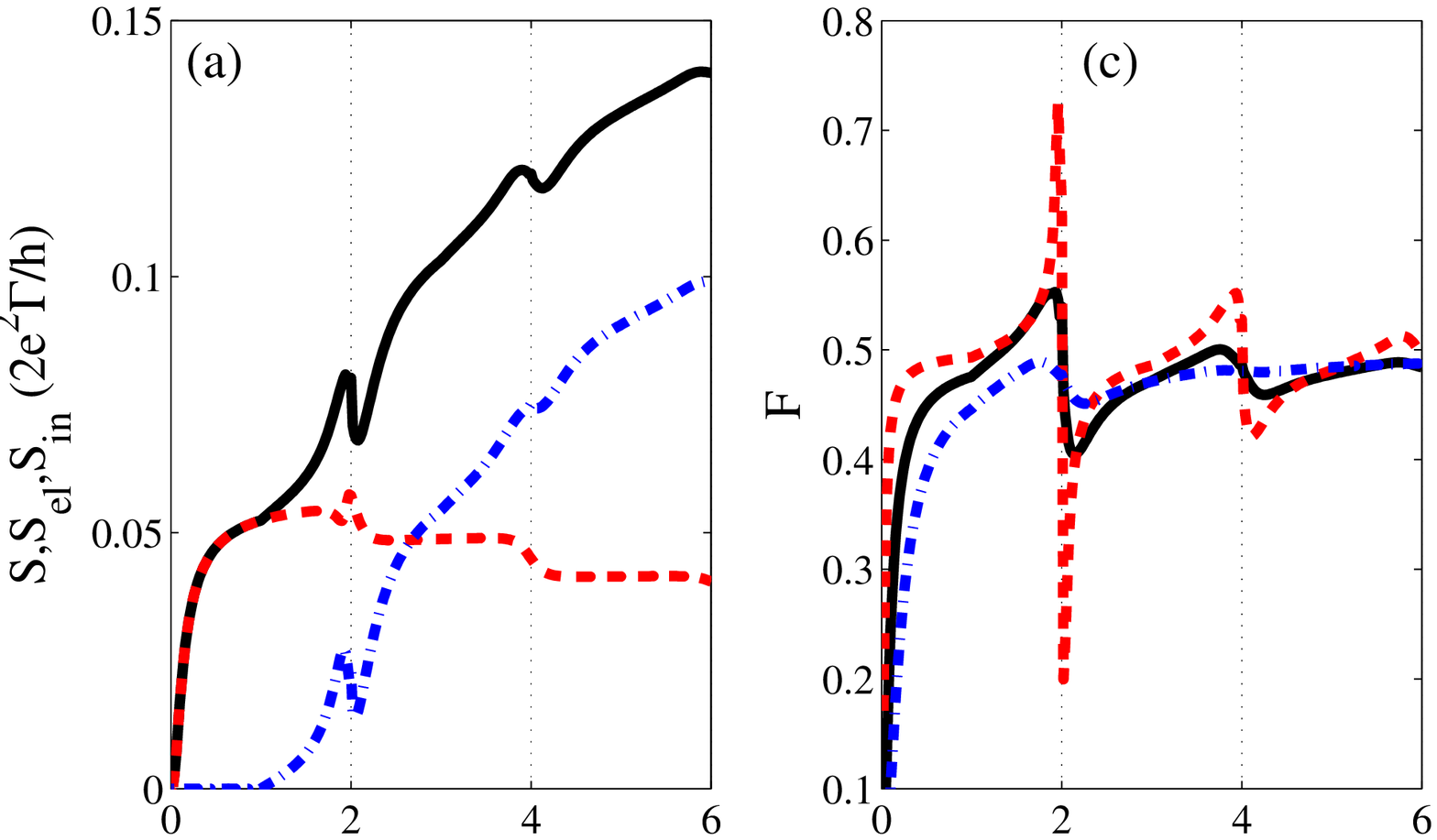}
\vspace{3mm}

\hspace*{-3mm}
\includegraphics[height=5.5cm,width=8.5cm]{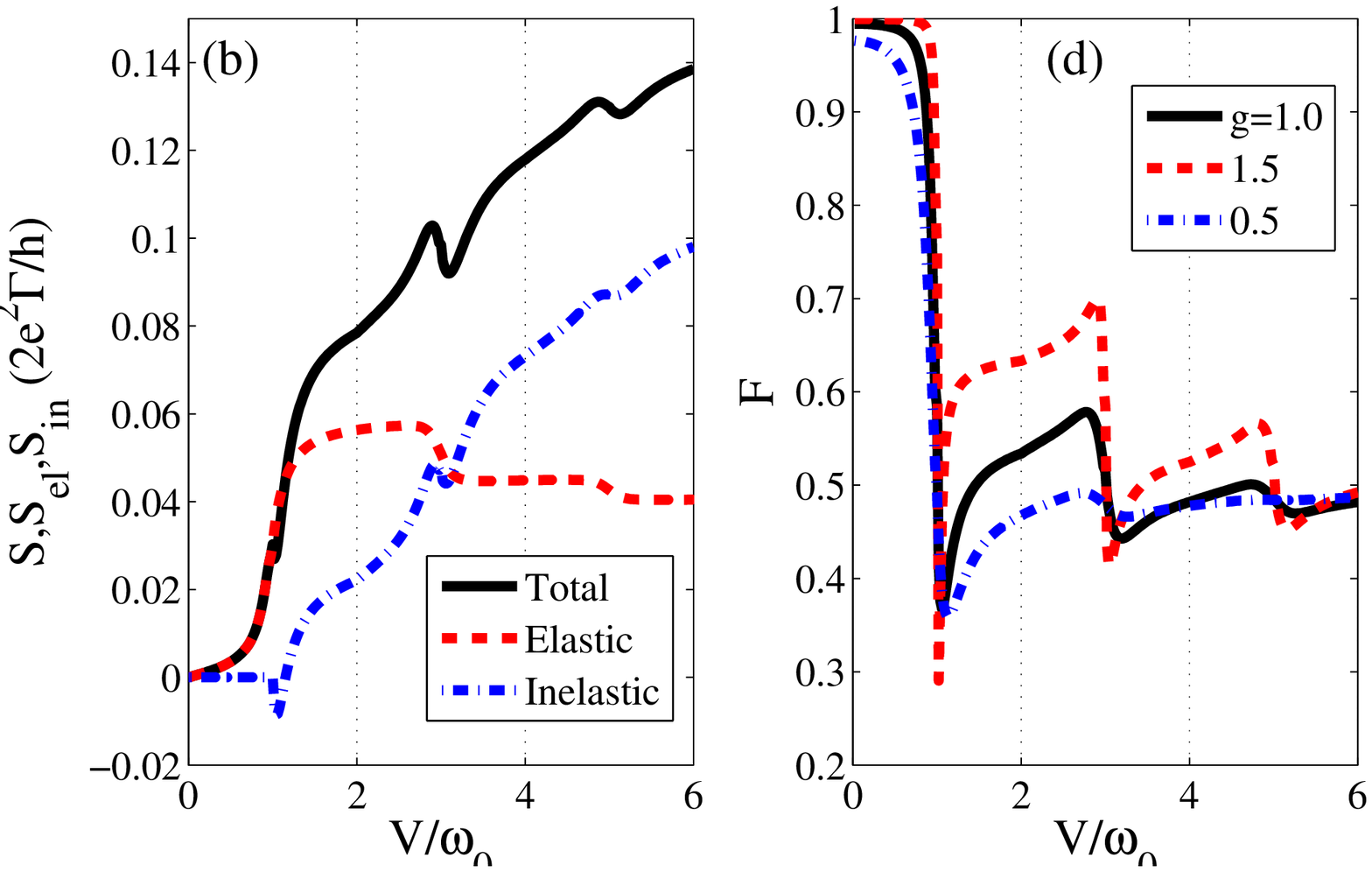}
\caption{(Colour online) (a, b) The zero-temperature shot noise (solid line), and its elastic (dashed line) and inelastic 
(dotted-dashed line) parts as functions of bias voltage for a single-molecular QD with $\widetilde{\varepsilon}_d=0$ (a) and 
$0.5\omega_0$ (b), respectively. The electron-phonon coupling constant is set to be $g=1.0$; (c, d) The corresponding Fano 
factors for the two systems, $\widetilde{\varepsilon}_d=0$ (c) and $0.5\omega_0$ (d), with different electron-phonon coupling 
constants $g=1.0$ (solid line), $1.5$ (dashed line), and $0.5$ (dotted-dashed line). We set $\Gamma=0.1\omega_0$ in the 
calculations.}
\label{fig6}
\end{figure}

In Figs.~\ref{fig6}(a) and (c), we plot the calculated shot noise and its two contributive parts as functions of bias voltage 
$V>0$ for the systems with $\widetilde{\varepsilon}_d=0$ and $\widetilde{\varepsilon}_d=0.5\omega_0$, respectively. It is 
observed that for the empty QD ($\widetilde{\varepsilon}_d=0.5\omega_0$), the shot noise of the elastic channel inherits the same 
behavior as the elastic current with increasing bias voltage, continuous increase up to the resonant point and  downward steps at 
$V=3\omega_0$ and $5\omega_0$. On the contrary, the inelastic shot noise exhibits abrupt downward jumps at $V=\omega_0$ and 
$3\omega_0$, instead of upward jumps in the inelastic current. At $V=\omega_0$, we can evaluate approximately the correction to 
the shot noise due to the inelastic tunneling for the system with $\widetilde{\varepsilon}_d=0.5\omega_0$ as
\bn
S_{in} &\simeq& 2e I_{in} - \frac{4e^2}{h}(\Gamma_L \Gamma_R)^2 w_0^2 w_1 (2w_0+w_1) \cr
&& \times \int_{\omega_0- V/2}^{V/2} d\omega \left [\frac{1}{{\cal D}_0^2(\omega)} + \frac{1}{{\cal D}_0^2(-\omega)} \right ] \cr
&\simeq& \frac{4e^2}{h} \frac{\Gamma_L \Gamma_R w_0 w_1}{|\Sigma_{ci}^r(0)|^2} \frac{(\Gamma_L-\Gamma_R)^2 - 4\Gamma_L \Gamma_R 
(1+ \frac{w_1}{w_0})}{(\Gamma_L + \Gamma_R)^2}. \nonumber
\en
For the symmetric tunnel-coupling case considered in this paper, $\Gamma_L=\Gamma_R$, the opening of inelastic channel generates 
a negative contribution to the shot noise. The same corrections of the inelastic noise will be predicted at $V=3\omega_0$ and 
$5\omega_0$, leading to downward jumps in the shot noise in association with the elastic noise. While the situation is more 
complex for the partially filled QD ($\widetilde{\varepsilon}_d=0$). At first, the inelastic noise shows a tiny upward jump at 
$V=\omega_0$, i.e. a positive correction, because of  
\bn
S_{in} &\simeq& \frac{4e^2}{h} \frac{\Gamma_L \Gamma_R w_0 w_1}{\left [ (\omega_0/2)^2+|\Sigma_{ci}^r(\omega_0/2)|^2 \right ]^2} 
\cr
&& \times \left [ (\omega_0/2)^2 + |\Sigma_{ci}^r(\omega_0/2)|^2 - \Gamma_L \Gamma_R w_0^2 (2+ \frac{w_1}{w_0}) \right ]. 
\nonumber
\en
But the inelastic noise becomes downward jump at $V=2\omega_0$ again.
Actually, the inelastic noise contribution has been examined for a QD with weak EPI, a soft phonon $\omega_0\ll \Gamma$, and 
arbitrary transmission based on the second-order perturbative calculation at $V=\omega_0$ where the inelastic channel is just 
opening.\cite{Schmidt} A sign change in the inelastic noise correction at certain domains in parameter space of transmission and 
energy level has been addressed and ascribed to the underlying competition between elastic and inelastic processes. Very 
recently, the negative contribution to noise has been experimentally observed on Au nanowires in the weak EPI limit and has been 
ascribed to the coherent two-electron tunneling processes assisted by phonon emission that reduce electronic fluctuations due to 
Pauli principle.\cite{Kumar}  
The present investigation in this paper indicates indeed that the interplay of elastic and inelastic scattering processes causes 
the following properties of shot noise: (1) the elastic shot noise exhibits a downward step at the bias voltages 
$V=2(n\omega_0\pm \widetilde{\varepsilon}_d)>0$ as the elastic current does; (2) Meanwhile, the opening of inelastic channel at 
these bias voltages induces an abrupt increase of the transmission probability of the inelastic channel [i.e. the inelastic 
current as shown in the above subsection, Figs.~\ref{fig4}(a) and (b)] and consequentially results in an obvious downward jump; 
otherwise the inelastic noise shows only a tiny increase; (3) In particular, for the molecular QD with an energy level of 
$\widetilde{\varepsilon}_d=0.5\omega_0$, the inelastic noise becomes negative at 
$V=2(\omega_0-\widetilde{\varepsilon}_d)=\omega_0$, i.e. the sign change of the inelastic correction to shot noise in the case of 
strong EPI and a hard phonon. Nevertheless, no such negative correction to noise was found in the PTA calculations.\cite{Maier} 
We argue that this is because the PTA considers only the elastic scattering processes as pointed out above.        

To analyze the relative strength of noise, a more useful quantity is the so-called Fano factor $F$ defined as the ration of the 
shot noise to the Poisson value, $F=S/2eI$. It is obvious from Eq.~(\ref{s0}) that the present approximation exhibits no 
super-Poissonian noise, being in agreement with the previous NGF calculation in Ref.~\onlinecite{Maier} under the same 
approximation, strong electron-phonon interaction and thermal equilibrated phonon. It also deserves to point out that the present 
result is in no conflict with that of the rate-equation calculations. Even though a giant Fano factor has been predicted due to 
avalanchelike transport of electrons by rate-equation calculations,\cite{Mitra,Koch} it has been subsequently clarified that a 
single-level molecular QD will exhibit super-Poissonian noise only when both of two conditions, external-bias-voltage-driven 
unequilibriated phonon and asymmetric tunnel-couplings between the QD and two leads, are simultaneously 
satisfied.\cite{Zazunov,Shen} Otherwise, the shot noise will decrease with increasing strength of dissipation of the hot phonon 
to environment, and eventually become sub-Poissonian noise and show steplike behavior.\cite{Shen,Dong2} In this paper, our NGF 
calculations predict more rich oscillatory behavior of the Fano factor as a function of the bias voltage, as shown in 
Figs.~\ref{fig6}(c) and (d). 
It is interesting to observe that the aforementioned downward jumps in the shot noise in conjunction with the upward steps in the 
current induce obvious dips in the Fano factor, whose values can be smaller than $1/2$. Since for a resonant tunneling model with 
a small tunneling rate $\Gamma$, the typical value of the Fano factor of a symmetric tunnel-junction at large bias voltage is 
right equal to $1/2$, this unusual smaller-than-one-half Fano factor therefore can be regarded as an unambiguous signature of 
{\em vibronic} participation in electronic tunneling.

\section{Conclusion}

In conclusion, in this paper we have investigated inelastic effects on the FCS of electronic tunneling through a single-molecular 
QD in the presence of strong electron-phonon interaction, weak tunnel-couplings, and hard phonon mode. For this purpose, we have 
performed the Lang-Firsov canonical transformation for the local electron-phonon interaction and made use of the non-crossing 
approximation to decouple the electronic and phononic degrees of freedom. Then we have employed the generalized nonequilbirium 
Green function technique for the FCS and derived an explicit analytical Levitov-Lesovik formula for the cumulant generating 
function under the approximation that the the molecular vibration is assumed to be always thermally equilibrated due to fast 
dissipation to a thermal phonon bath, i.e. the environment. This formula can not only provide fundamental knowledge of how to 
clarify independent elementary processes in the vibration-assisted charge transfer, but also give analytical expressions for the 
tunneling current and its zero-frequency shot noise. Subsequently, we have carried out numerical calculations for the current and 
shot noise of a QD with symmetric tunnel-couplings at zero temperature and further analyzed their bias-voltage dependence in 
detail. 

Even though several of our formal results, for example, the upward or downward jumps in the current and shot noise only at 
$V=\omega_0$, were already addressed in previous papers by the second-order perturbative calculations for weak EPI 
system,\cite{Schmidt,Mitra,Egger,Entin} there are still some debates in these issues in the literature. The present paper has 
provided complementary investigation for strong EPI system. We have found that: (i) The singularities in the electronic 
self-energy and spectral function cause discontinuities in the zero-frequency shot noise in the weak tunnel-coupling case, i.e. 
weak bare elastic transparency of the molecular junction. The sign of the discontinuity occurring at $V=\omega_0$ (single-phonon 
scattering process) depends on the normalized energy level of the molecular QD. For an empty QD, 
$\widetilde{\varepsilon}_d=0.5\omega_0$, the inelastic channel provides a negative contribution to noise at $V=\omega_0$; 
otherwise, a positive contribution is observed. Moreover, multi-phonon scattering events will always induce downward jumps. It is 
noticed that the opening of inelastic channel can also affect the elastic channel, leading to downward steps in the elastic part 
of the current and shot noise; (ii) Contrary to the results of rate-equation calculations, our investigations predict oscillatory 
structure and apparent dips in the Fano factor. The small Fano factor, $F<1/2$, can be considered as a typical characteristics of 
phonon-assisted electronic tunneling through a single molecular junction.    

Noticeably, our approximative calculations for the strong EPI system with an equilibrated phonon have reproduced the logarithmic 
singularities in the electronic self-energies\cite{Engelsberg} and consequently found the discontinuities in the differential 
conductance and shot noise. It is therefore desirable in the future research to develop a fully self-consistent calculation, i.e. 
solving the coupled Dyson equations for the electronic GF $G_c(t,t')$ and the phononic GF $K(t,t')$ simultaneously, to observe 
the unequilibrated phonon effect on the singularities and discontinuities.

\begin{acknowledgments}

This work was supported by Projects of the National Basic Research Program of China (973 Program) under Grant No. 2011CB925603, 
and the National Science Foundation of China, Specialized Research Fund for the Doctoral Program of Higher Education (SRFDP) of 
China. 

\end{acknowledgments}

\end{document}